\documentclass[aps, prl, twocolumn]{revtex4-1}
\usepackage{graphicx}
\usepackage{subfigure}
\usepackage{amsmath}
\usepackage[bookmarks,hyperindex]{hyperref}
\usepackage{listings}
\newcommand{\subfigimg}[3][,]{%
  \setbox1=\hbox{\includegraphics[#1]{#3}}
  \leavevmode\rlap{\usebox1}
  \rlap{\hspace*{40pt}\raisebox{\dimexpr\ht1-2\baselineskip}{#2}}
  \phantom{\usebox1}
}
\makeatletter
\begin{document}
\title{Computation of the correlated metal-insulator transition in vanadium dioxide from first principles}
\author{Huihuo Zheng and Lucas K. Wagner}
\email{lkwagner@illinois.edu}
\affiliation{Department of Physics, University of Illinois at Urbana-Champaign Urbana, IL 61801-3080, USA}

\begin{abstract} 
Vanadium dioxide(VO$_2$) is a paradigmatic example of a strongly correlated system that undergoes a
metal-insulator transition at a structural phase transition. 
To date, this transition has necessitated significant post-hoc adjustments to theory in order to be described properly.
Here we report standard state-of-the-art first principles quantum Monte Carlo (QMC) calculations of the structural dependence of the properties of VO$_2$. 
Using this technique, we simulate the interactions between electrons explicitly, which allows for the metal-insulator transition to naturally emerge, importantly without {\it ad-hoc} adjustments.
The QMC calculations show that the structural transition directly causes the metal-insulator transition and a change in the coupling of vanadium spins.
This change in the spin coupling results in a prediction of a momentum-independent magnetic excitation in the insulating state.
While two-body correlations are important to set the stage for this transition, they do not change significantly when VO$_2$ becomes an insulator.
These results show that it is now possible to account for electron correlations in a quantitatively accurate way that is also specific to materials.
\end{abstract}

\maketitle
Systems of strongly correlated electrons at the border between a metal-insulator transition
can result in a variety of unique and technologically useful behavior, such as
high-temperature superconductivity\cite{HTC1990} and collossal
magnetoresistance\cite{Colossal1997}. In vanadium dioxide (VO$_2$), the metal-insulator transition (MIT)
occurs at $T=340$K \cite{Morin1959}, at which the conductivity decreases by more than 4 orders
of magnitude. This MIT is accompanied by a structural change from rutile (P4$_2$/mnm) to
monoclinic (P2$_1$/c) \cite{Andersson1956}, as well as a transition in the magnetic
susceptibility from a paramagnet-like Curie-Wiess law to a temperature-independent form. In
the rutile phase, the vanadium atoms are located at the centers of octahedra formed by the
oxygen atoms; chains of equidistant vanadium atoms lie along the c axis ([001]). In the
low-temperature monoclinic phase, vanadium atoms shift from the centers of the oxygen
octahedra and form a zig-zag pattern consisting of V dimers. It is a long-standing question
whether the MIT is primarily caused by the structural change that doubles the unit cell
(Peierls distortion), or by correlation effects that drive the system to become insulating
\cite{Goodenough1960, Mott1975,LDADMFT2012}, or potentially some mixture of the two.

The metal-insulator transition in VO$_2$ is unusually challenging to describe.
Standard
density functional theory (DFT) \cite{Eyert2002,Eyert2011} obtains metallic states for both
structures, while corrections based on an effective Hubbard $U$\cite{Eyert2011,LDAU2009} or
hybrid functionals\cite{WhyHSE2012} often obtain insulating states for both structures.
Similarly, DFT+DMFT\cite{LDADMFT2004, DMFTSinglet, LDADMFT2012} and
DFT+GW\cite{Continenza1999, Gatti2007,Sakuma2009} calculations indicate that how the
correlation is treated changes the calculated properties dramatically. 
Since there is no {\em a-priori} guideline for these theories, while they are descriptive and valuable techniques, their reliability is uncertain in a predictive capacity.
This issue is a severe constraint for the design and study of correlated electron systems.

\begin{figure}
   \centering
   \includegraphics[width=0.40\textwidth, clip]{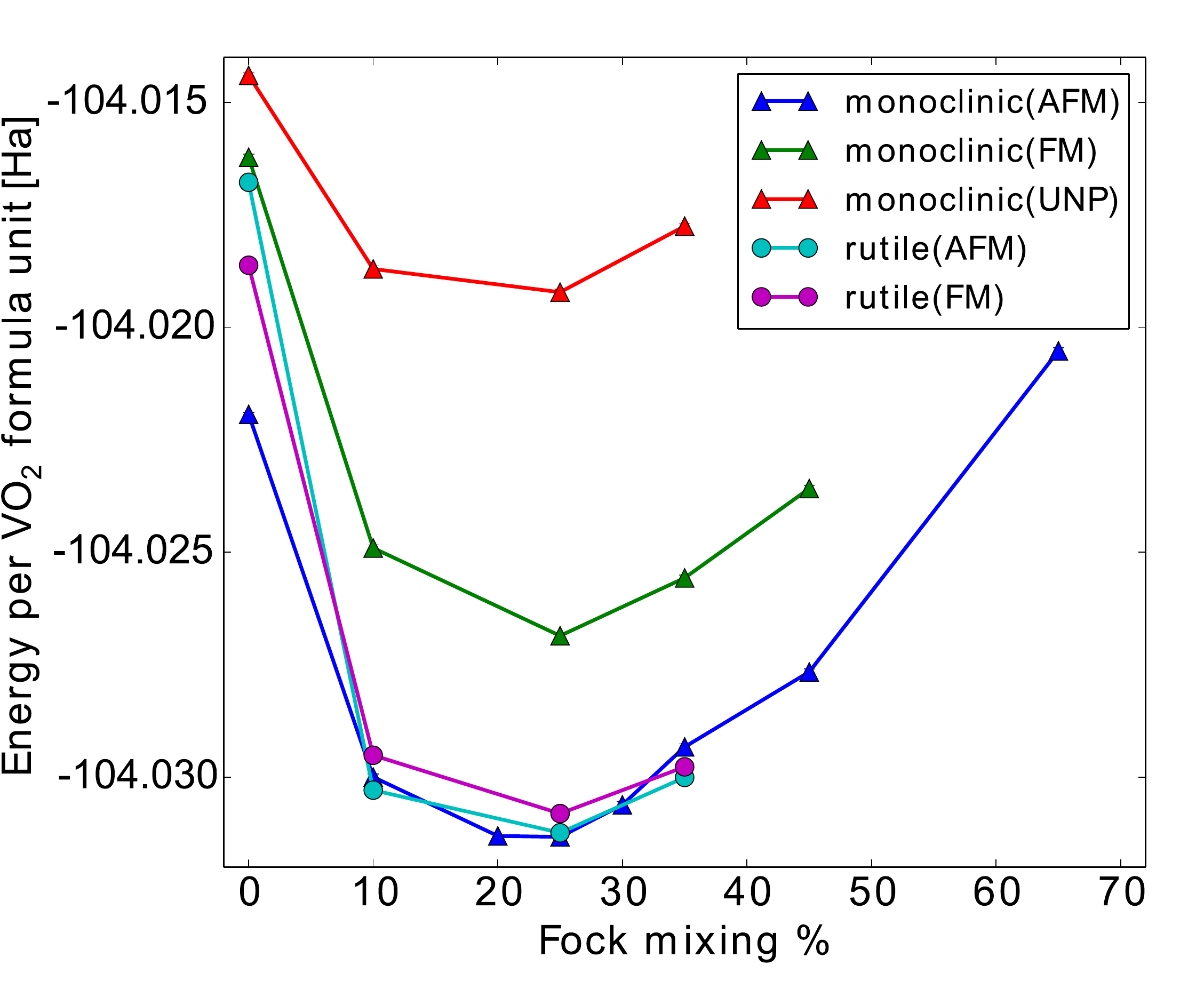}
   \caption{FN-DMC energies of VO$_2$ with trial wave functions from DFT of different hybrid
functionals. }\label{fig:nodal}
\end{figure}

In this study, we use the explicitly correlated fixed-node diffusion quantum Monte Carlo (FN-DMC) \cite{QMC} to
investigate the electronic structures of the rutile and monoclinic VO$_2$ from first
principles. FN-DMC has been shown to be a highly accurate method on other transition metal
oxides\cite{Kolorenc2008,wagner_effect_2014}. In this method, one explicitly samples
many-body electronic configurations using Coulomb's law for interactions, which allows for
the description of correlation effects without effective parameters. 
We show that FN-DMC correctly characterizes the electronic structure and magnetic response of VO$_2$ in the two phases. 
Our calculations provide quantitative microscopic details of the structure-dependent 
spin couplings between vanadium atoms. It clearly reveals how the structural distortion changes the interatomic 
hybridization between vanadium and oxygen, which results in a significant change of the superexchange magnetic coupling between
vanadium atoms. Monoclinic VO$_2$ is in a non-magnetic singlet state consisting of spin dimers due to strong intradimer coupling. 
The calculations contain a singlet-triplet spin excitation of 123(6) eV in monoclinic VO$_2$, which can be verified in experiment to test the predictive power of this method.

\begin{figure}
  \centering
  \begin{tabular}{@{}p{0.95\linewidth}@{\quad}p{0.95\linewidth}@{}}
    \subfigimg[width=\linewidth, clip]{(a)}{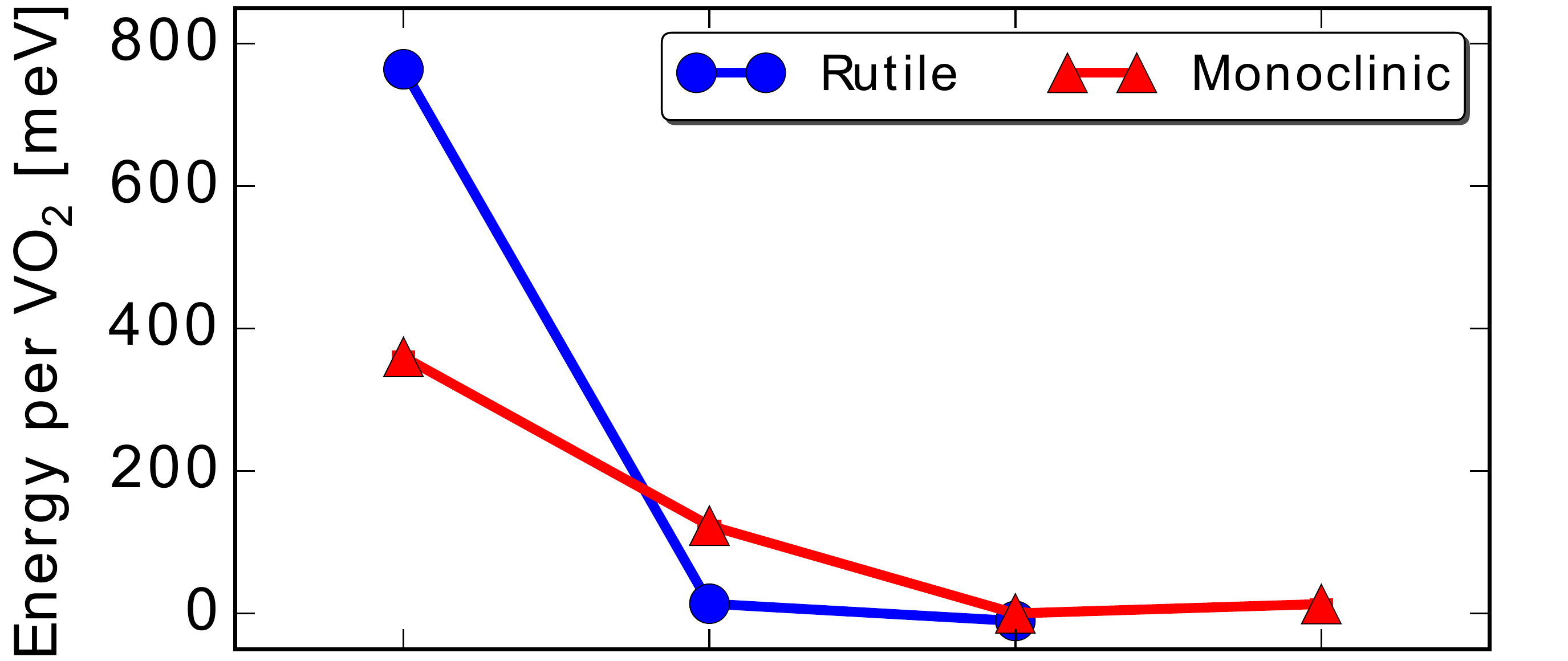} \\\vspace*{-1.2em}
    \subfigimg[width=\linewidth, clip]{(b)}{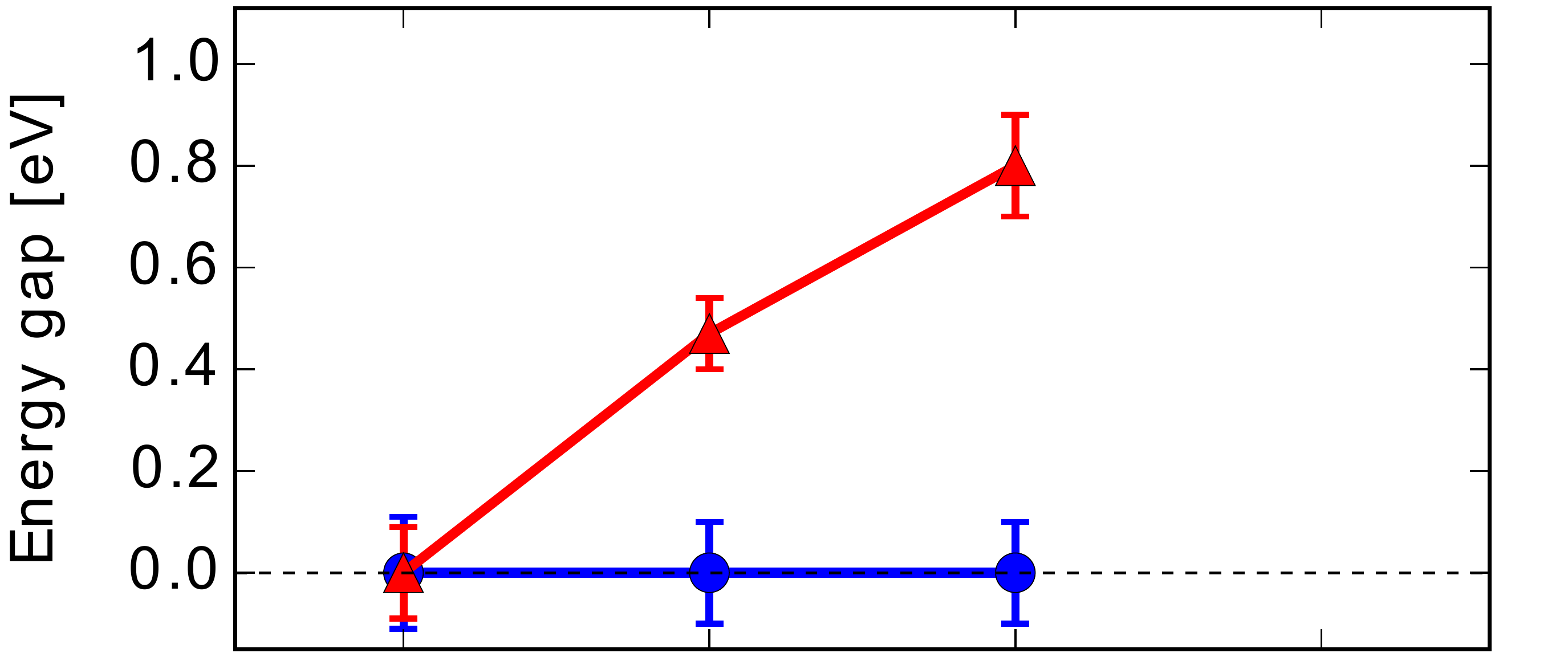} \\\vspace*{-1.2em}
    \subfigimg[width=\linewidth, clip]{(c)}{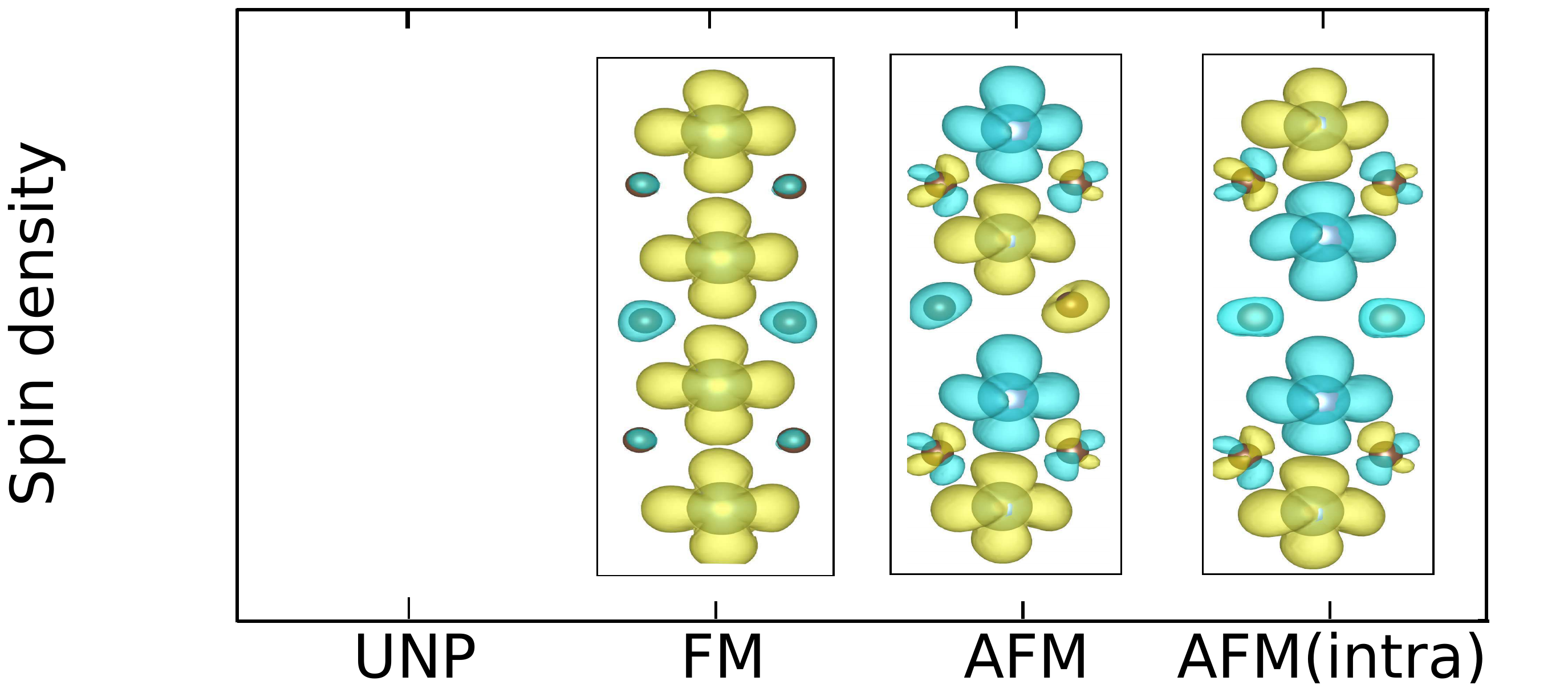} 
  \end{tabular}
\caption{(Color online) FN-DMC energetic results: (a) per VO$_2$ unit for different magnetically ordered states, spin-unpolarized, ferromagnetic (FM), and antiferromagnetic [AFM and
AFM (intra)]. Energies are referenced to AFM monoclinic VO$_2$.(b) Optical gaps for various states. (c) Spin density for various states.}
\label{fig:energy}
\end{figure}

\begin{figure}
   \centering
   \includegraphics[width=0.40\textwidth, clip]{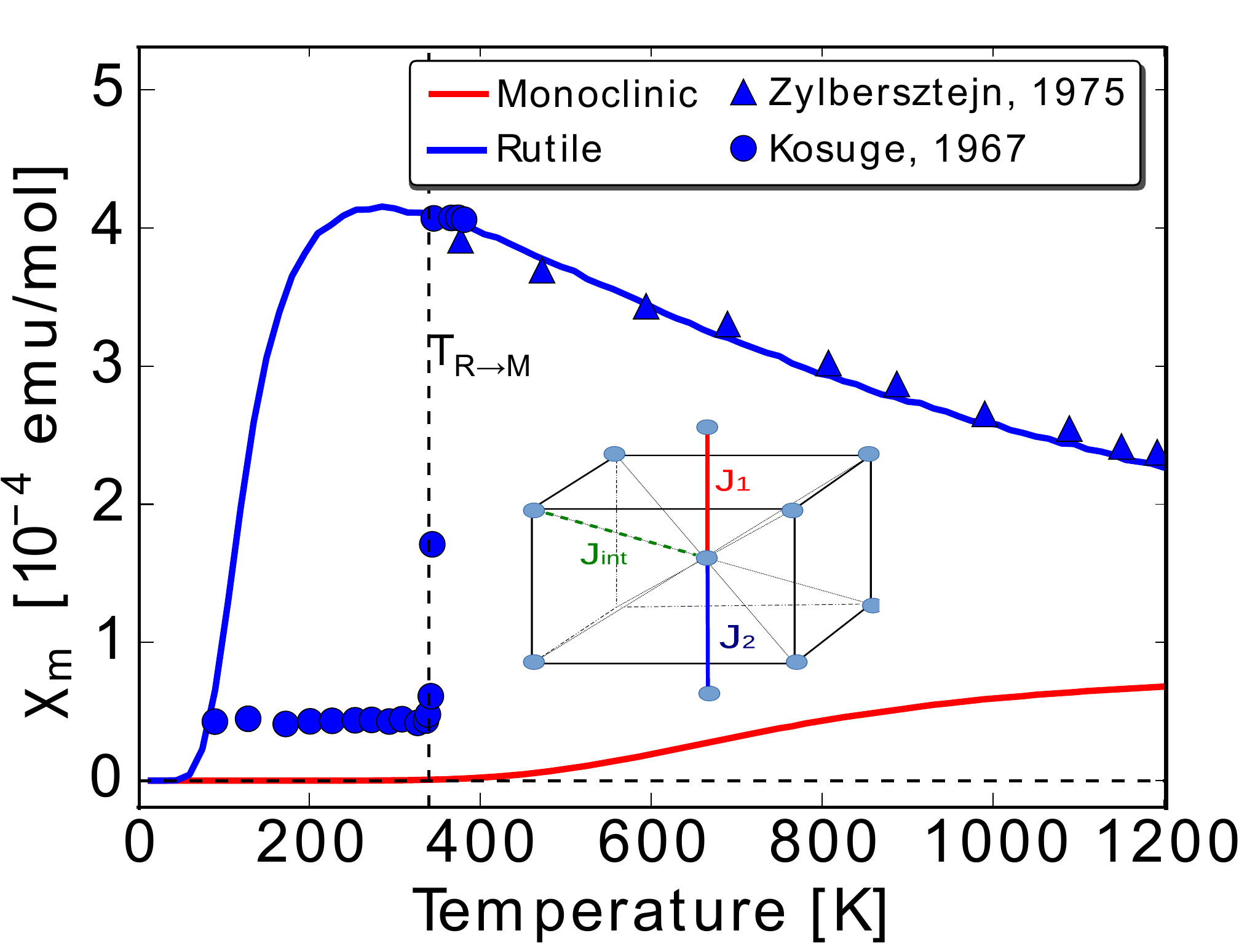}
   \caption{Temperature dependence of magnetic susceptibility obtained by Ising model
   simulation on VO$_2$ lattice, with magnetic coupling from FN-DMC compared to results from Zylbersztejn\cite{Mott1975} and Kosuge\cite{Mossbauer1967}. There has been an overall scale applied on the data, and the transition temperature is indicated by the verticle dashed line. }
\label{fig:ising}
\end{figure}

The calculations were performed as follows. The crystal structure of the rutile and
monoclinic phases were taken from experiment \cite{Ishiwata2010}. DFT calculations were
performed using the CRYSTAL package, with initial spin configurations set to aligned,
anti-aligned, or unpolarized vanadium atoms. 
Different exchange-correlation functionals with varying levels of Hartree-Fock
exchange were used: 
\begin{eqnarray}\label{eq:exc} 
E_{\text xc}=(1-p)E_x^{PBE} + p E_x^{HF} +
E_c^{PBE} 
\end{eqnarray} 
where $E_x^{PBE}$ and $E_c^{PBE}$ are the Perdew-Burke-Ernzerhof
(PBE) exchange and correlation functionals respectively. The simulations were performed on a
supercell including 16 VO$_2$ formula units with 400 valence electrons. 
An $8\times 4\times 4$ Monkhorst-Pack k-grid was chosen for sampling the
first Brillouin zone of the simulation cell. A Burkatzki-Filippi-Dolg (BFD) pseudopotential
\cite{BFD2007, BFD2008} was used to represent the He core in oxygen and Ne core in vanadium.
The band structure obtained using the BFD pseudopotential shows good agreement with the
all-electron DFT calculation \cite{EXCITING} (using PBE functional). 

The result of the DFT calculations is a set of Slater determinants made of Kohn-Sham
orbitals that have varying transition metal-oxygen hybridization and spin orders. A Jastrow
correlation factor was then added to these Slater determinants as the trial wave functions
for quantum Monte Carlo calculations \cite{Qwalk}. Total energies were averaged over twisted
boundary conditions and finite size errors were checked to ensure that they are negligible
(see Supplementary Information). The fixed-node error in FN-DMC was checked by comparing the
energetic results from different trial wave functions from DFT calculations with different
$p$ in Eqn~\ref{eq:exc}. The trial wave functions corresponding to the 25\% Hartree-Fock
mixing (PBE0 functional \cite{PBE0}) produce the minimum FN-DMC energy (Fig~\ref{fig:nodal}).
The behavior in VO$_2$ is commonly seen in other transition metal oxides
\cite{Kolorenc2010}. Thus, all our FN-DMC results in the main manuscript were produced using 25\%
mixing. The gap was determined by promoting an electron from the highest occupied band to
the lowest unoccupied orbital in the Slater determinant, then using that determinant as a
trial function for FN-DMC. 

The energetic results of the quantum Monte Carlo calculations are summarized in
Fig~\ref{fig:energy}. Both the rutile and monoclinic structures have lowest energy with
antiferromagnetic ordering of the spins. The unpolarized trial function has sufficiently
high energy to remove it from consideration of the low energy physics. The energy difference
between the ferromagnetic and antiferromagnetic orderings changes from 24(6) meV to 123(6)
meV from the rutile to monoclinic structure. The energy difference between the lowest energy
spin orderings for rutile and monoclinic is 10(6) meV, which is within statistical
uncertainty of zero. The latent heat is 44.2(3) meV\cite{Berglund1969}; the small
descrepancy may be due to either finite temperature or nuclear quantum effects, or fixed
node error. In the monoclinic structure, the vanadium atoms are dimerized, which allows for
a type of magnetic ordering in which the intra-dimer vanadium dimers are aligned. This
ordering increases the energy by 13(6) meV.

The lowest energy wave functions all have magnetic moments on the vanadium atoms close to 1 Bohr magneton.
In the rutile structure, the spins are coupled with a small superexchange energy along the $c$ axis.
In the monoclinic structure, the spins are coupled strongly within the vanadium dimers and weakly between them.
The spin coupling within the dimers should give rise to a spin excitation with little dispersion at approximately 123(6) meV, which could potentially be observed with neutron spectroscopy.
This excitation has been proposed in the past by Mott\cite{mott_metal-insulator_1974}; our results here provide a precise number for this excitation.

\begin{figure*}
\centering
\includegraphics[width=\textwidth]{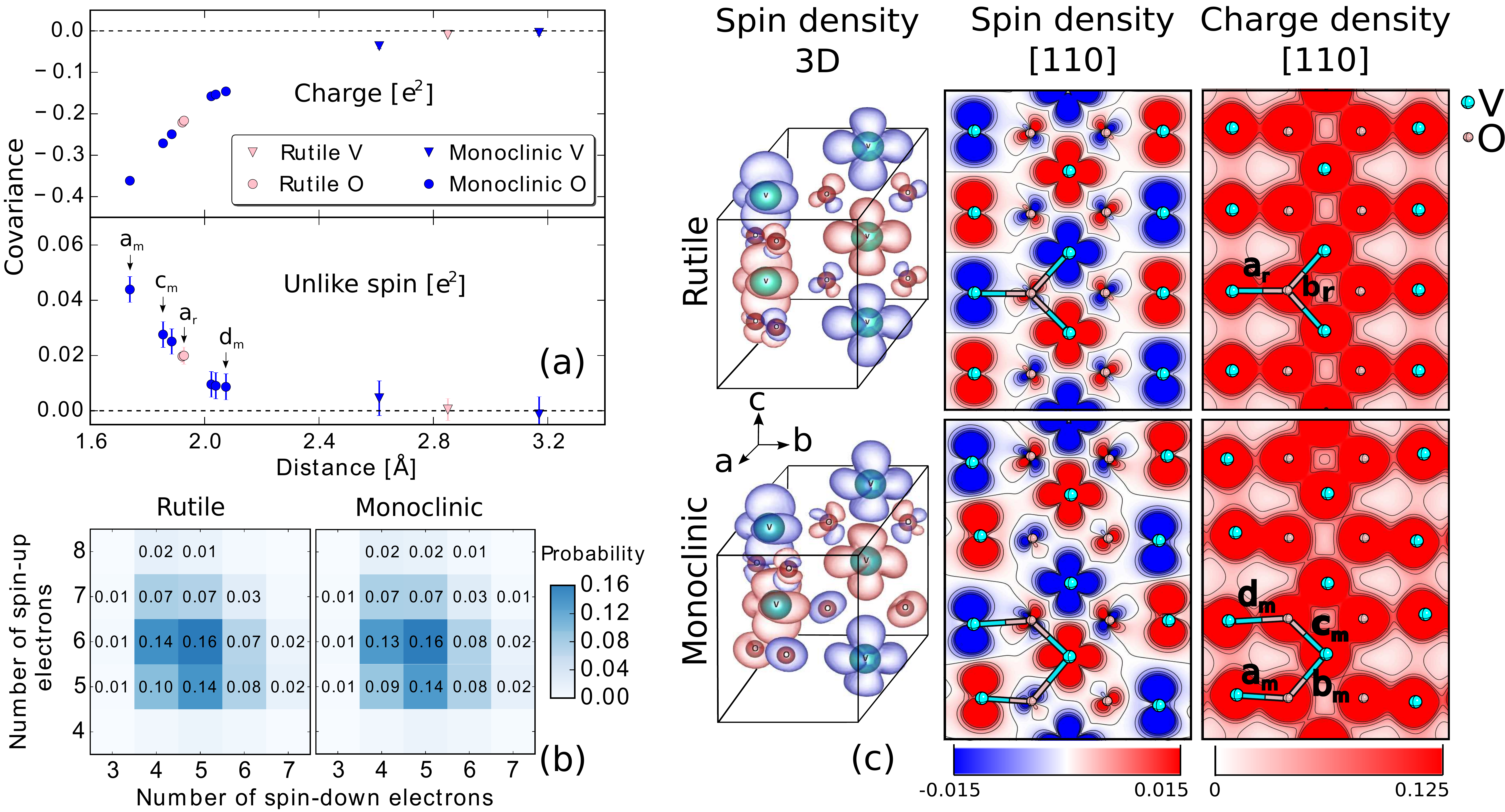}
\caption{(Color online) Change of V-O hybridizations manifested through: (a) Inter-site charge ($n_i$), and unlike-spin ($n_{i\uparrow}$ or
$n_{i\downarrow}$) covariance quantified as $\langle O_i O_j\rangle-\langle O_i\rangle
\langle O_j\rangle$, where $O_i$ is the onsite value of specific physical quantities. The inter-site covariance between a chosen vanadium center and the surrounding atoms is plotted as a function of the inter-atomic distance. (b) Onsite spin-resolved probability distribution on vanadium atoms -- $\rho(n_{i}^{\uparrow}, n_{i}^{\downarrow})$. (c) Spin and charge density of the rutile and monoclinic VO$_2$. Figures on the left panel are 3D isosurface plots of spin density, and on the right panel are contour plots of spin and charge density on the $[110]$ plane. }
\label{fig:covariance}
\end{figure*}

To study whether the magnetic behavior from the energies in Fig~\ref{fig:energy} are
consistent with experiment, we make a simple Ising model.  In our model, the spins are on
vanadium sites, and we only consider couplings between adjacent sites: (1)$J_1$ --
intra-dimer coupling; (2) $J_2$ -- inter-dimer coupling; (3) $J_{\text int}$ -- nearest
inter-chain coupling (see Fig.~\ref{fig:ising}). We assume that the energy for a magnetic state 
takes the following form, 
\begin{eqnarray}
E&=&J_1\sum_{\text{intradimer}<i,j>}\sigma_i\sigma_j + J_2\sum_{\text{interdimer}<i,j>} \sigma_i\sigma_j \nonumber\\
 &&+ J_{\text int}\sum_{\text{interchain} <i,j>}\sigma_i \sigma_j + E_0\,.
\end{eqnarray} 
We fit $J_1$ and $J_2$ to the energetic results of the FM, AFM, and AFM(intra) orderings.
For the monoclinic structure, $J_1 = 123$ meV, $J_2=13$ meV, while for the rutile structure  $J_1 = J_2=10$ meV.
Since we did not compute the
inter-chain coupling strength $J_{\text
int}$, we set it to be $2.5$ meV which is reasonably small compared to the intra-chain coupling. 
The result is not sensitive to the inter-chain coupling as long as it is small. 
We then perform a Monte Carlo simulation is on a $10\times 10\times 10 $ super
cell, in which the finite size effect has been checked to be small. The results are presented in
Fig~\ref{fig:ising}. The Curie-Weiss behavior of the magnetic susceptibility above the
transition temperature is reproduced very well, while the flat susceptibility below the
transition is also reproduced.

Moving to the gap properties (Fig~\ref{fig:energy}b), the gap of the low-energy AFM ordering in the rutile structure is zero, 
while the gap in the monoclinic is 0.8(1) eV. 
This compares favorably to experiment, which have gaps of zero and 0.6--0.7 eV\cite{ExpGap}.
Meanwhile, in the higher energy FM ordering, the monoclinic structure has a minimal gap of 0.42(7) eV and the rutile structure has a gap of 0.0(1) eV, both in the spin-majority channel.
If there are not unpaired electrons on the vanadium atoms, then the gap is zero. 


From the above energy considerations, a few things become clear about the results of the FN-DMC calculation.
The gap formation is not due to a particular spin orientation, so the transition is not of Slater type.
However, it is dependent on the formation of unpaired electrons on the vanadium atoms.
These behaviors might lead one to suspect that the transition is of Mott type, since the gap in the monoclinic structure is a $d\rightarrow d$ transition.
In the classic Mott-Hubbard model, the metal-insulator transition is a function of $U/t$, where $U$ is the on-site repulsion and $t$ is the site-to-site hopping.

To make the physics more clear, we connect the detailed quantum results to an approximate low-energy Hubbard-like model.
We define the V sites and O sites by the Voronoi polyhedra surrounding the nuclei.
For a given sample in the FN-DMC calculation, we evaluate the number of up spins $n_i^\uparrow$ and the number of down spins $n_i^\downarrow$ on a given site $i$.
We then histogram the joint probability to obtain a set of functions $\rho_{i, j, \sigma_i, \sigma_j}(n_i^{\sigma_i},n_j^{\sigma_j})$, where $i,j$ are site indices and $\sigma_i,\sigma_j$ are spin indices.
The connection to $t$ and $U$ are made through covariances of these number operators.
$t$ is connected to the covariance in the total number of electrons on a site $i$, $n_i=n_i^\uparrow+n_i^\downarrow$ with site $j$: $\langle (n_i-\langle n_i \rangle)(n_j-\langle n_j \rangle) \rangle$. 
If this charge covariance is large, then the two sites share electrons and thus $t$ between those two sites is large.
$U/{\bar t}$, where ${\bar t}$ is the average hopping, is connected to the covariances in the number of up electrons and down electrons on a given site: $\langle (n_i^\uparrow-\langle n_i^\uparrow \rangle)(n_i^\downarrow-\langle n_i^\downarrow \rangle) \rangle$.
This quantity, which we term the onsite spin covariance, is zero for Slater determinants and is a measure of the correlation on a given site.

Fig~\ref{fig:covariance}a shows the covariances evaluated for the AFM ordering of the two structures of VO$_2$.
The most striking feature is that the charge covariance changes dramatically between the two structures.
This feature can also be seen in the charge density in Fig~\ref{fig:covariance}c.
The dimerization causes a large change in the hybridization between vanadium and oxygen atoms.
In particular, the largest hybridization is now not within the dimers, but between the vanadium atoms and the oxygen atoms in the adjacent chain, denoted $a_m$ in Fig~\ref{fig:covariance}b.
The intra-dimer hybridization is enhanced, and the inter-dimer hybridization is suppressed.
There are thus large rearrangements in the effective value of $t$, although the average value ${\bar t}$ is approximately constant.

On the other hand, the onsite unlike spin covariance does not change within stochastic uncertainties.
In Fig~\ref{fig:covariance}b, the joint probability function of $n_\uparrow$ and $n_\downarrow$ averaged over vanadium atoms with a net up spin are shown for both rutile and monoclinic phases. 
The correlations between up and down electrons are identical within statistical uncertainties in the two structures.
Therefore, $U/{\bar t}$ does not change very much between the two structures.
A one dimensitional Hubbard model would have $U/{\bar t} \simeq 2.0$ to obtain the unlike spin covariance that we observe, which puts VO$_2$ in the moderately correlated regime.

By performing detailed calculations of electron correlations within VO$_2$, we have shown that it is possible to describe the metal-insulator transition by simply changing the structure. 
To obtain the essential physics, it appears that the change in structure is enough to cause the metal-insulator transition.
As has been noted before\cite{Gatti2007}, the calculated properties of VO$_2$ are exceptionally sensitive to the way in which correlation is treated.
It is thus a detailed test of a method to describe this transition.
Fixed node diffusion quantum Monte Carlo passed this test with rather simple nodal surfaces, which is encouraging for future studies on correlated systems.
This method, historically relegated to studies of model systems and very simple {\em ab-initio} models, can now be applied to {\em ab-initio} models of correlated electron systems such as VO$_2$ and other Mott-like systems\cite{Kolorenc2008,wagner_effect_2014,foyevtsova_ab_2014}.

From the quantitatively accurate simulations of electron correlations, a simple qualitative picture arises. 
In both phases, there are net spins on the vanadium atoms with moderate electron-electron interactions compared to the hopping.
In the rutile phase, the vanadium oxide chains are intact with large hopping and small superexchange energy and thus the material is a correlated paramagnetic metal.
In the monoclinic phase, the dimerization reduces the interdimer hopping, primarily by interchain V-O coupling. The intradimer magnetic coupling increases because of an increase of intradimer V-O coupling. 
The spins then condense into dimers and a gap forms.
This can be viewed as a spin Peierls-like transition.

The results contained in this work, alongside other recent results show that the dream of simulating the many-body quantum problem for real materials to high accuracy is becoming achievable.
This accomplishment is a lynchpin for the success of computational design of correlated electron systems, since these calculations can achieve very high accuracy using only the positions of the atoms as input.
We have demonstrated that clear predictions for experiment can be made using {\it ab-initio} quantum Monte Carlo techniques, in particular the value of the singlet-triplet excitation in the spin-dimers of VO$_2$. 
If this prediction is verified, then it will be clear that these techniques can provide an important component to correlated electron systems design.

\section*{Acknowledgement}
The authors would like to thank David Ceperley for helpful discussions. This work was
supported by NSF DMR 12-06242 (LKW) and the Strategic Research Initiative at the University
of Illinois (HZ). The computation resources were from Blue Waters (PRAC-jmp), Taub (UIUC
NCSA), and Kraken (XSEDE Grant DMR 120042).
\bibliography{vo2}
\end{document}